\documentclass[structabstract]{aa}  

\usepackage[switch]{lineno}
\usepackage{graphicx}
\usepackage{subfig}
\usepackage{txfonts,float}
\usepackage{booktabs}
\usepackage{natbib}
\usepackage[utf8]{inputenc}
\usepackage{multirow}
\bibpunct{(}{)}{;}{a}{}{,} 
\usepackage{float}

\usepackage{longtable}

\newcommand{\swift}{\textit{Swift}}

\usepackage[colorlinks=true,linkcolor=red,citecolor=blue]{hyperref}
\usepackage[usenames,dvipsnames]{color}
\usepackage{colortbl,xcolor}

\begin{document}
\title{A quiescent galaxy at the position of the long GRB 050219A
\thanks{Based on observations
collected with GROND at the 2.2m telescope of the La Silla Observatory, Chile (PI: J. Greiner), at the Very Large Telescope of
the European Southern Observatory, Paranal, Chile (089.A-0843, PI: S. Piranomonte), 
and with the ESA space observatory {\it Herschel} (PI: L. Hunt)}}

   \author{
   A. Rossi\inst{1,2},
   S.~Piranomonte\inst{3},
   S.~Savaglio\inst{4,5,6},
   E.~Palazzi\inst{1},
   M.~J.~Micha{\l}owski\inst{7}, 
   S.~Klose\inst{2},
   L.~K.~Hunt\inst{8},
   L.~Amati\inst{1},
   J.~Elliott\inst{4},
   J.~Greiner\inst{4},
   C.~Guidorzi\inst{9},
   J.~Japelj\inst{10},
   D.~A.~Kann\inst{2}, 
   B.~Lo~Faro\inst{11},
   A.~Nicuesa Guelbenzu\inst{2},
   S.~Schulze\inst{12,13},
   S.~D.~Vergani\inst{14,15},
   L.~A.~Arnold\inst{16},
   S.~Covino\inst{15},
   V.~D'Elia\inst{17,3},
   P.~Ferrero\inst{18},
   R.~Filgas\inst{19}, 
   P.~Goldoni\inst{20},
   A.~K\"upc\"u~Yolda\c{s}\inst{21},
   D.~Le~Borgne\inst{22},
   E.~Pian\inst{1,23}, 
   P.~Schady\inst{4},
   G.~Stratta\inst{24}
}
  
\offprints{A. Rossi, a.rossi@iasfbo.inaf.it}

\institute{
   INAF-IASF Bologna, Area della Ricerca CNR, via Gobetti 101, I--40129 Bologna, Italy 
   \and
   Th\"uringer Landessternwarte Tautenburg, Sternwarte 5, 07778 Tautenburg, Germany 
   \and
   INAF-Osservatorio Astronomico di Roma, Via di Frascati 33, 00040 Monte Porzio Catone, Italy 
   \and
   Max-Planck-Institut f\"ur Extraterrestrische Physik, Giessenbachstra{\ss}e 1, 85748 Garching, Germany 
   \and
   Dipartimento di Fisica, Università della Calabria, 87036 Arcavacata di Rende, Italy  
   \and
   European Southern Observatory, 85748 Garching, Germany 
   \and
   SUPA, Institute for Astronomy, University of Edinburgh, Royal Observatory, Edinburgh, EH9 3HJ, UK 
   \and
   INAF-Osservatorio Astrofisico di Arcetri, Largo Fermi 5, I-50125 Firenze, Italy 
   \and
   Department of Physics and Earth Sciences, University of Ferrara, via Saragat 1, I-44122 Ferrara, Italy 
   \and
   Faculty of Mathematics and Physics, University of Ljubljana, Jadranska ulica 19, SI-1000 Ljubljana, Slovenia 
   \and
   Aix-Marseille Universit\`{e}, CNRS, LAM (Laboratoire d'Astrophysique de Marseille) UMR7326, 13388, France 
   \and
   Instituto de Astrof\'isica, Pontificia Universidad Cat\'olica, Av. Vicu\~na Mackenna 4860, Santiago, Chile 
   \and
   Millennium Institute of Astrophysics 
   \and
   GEPI, Observatoire de Paris, CNRS, Univ. Paris Diderot, 5 Place Jules Jannsen, F-92195, Meudon, France 
   \and
   INAF - Osservatorio Astronomico di Brera , via E. Bianchi 46, I-23807, Merate (LC), Italy 
   \and
   University of Rochester, Department of Physics and Astronomy, Rochester, NY 14627, USA 
   \and
   ASI, Science Data Centre, Via del Politecnico snc, I-00133 Roma, Italy 
   \and   
   Instituto de Astrofísica de Andalucía (IAA-CSIC), Glorieta de la Astronomía s/n, E-1008, Granada, Spain  
   \and
   Institute of Experimental and Applied Physics, Czech Technical University in Prague, Horska 3a/22, 12800 Prague 2, Czech Republic 
   \and
   APC, Univ Paris Diderot, CNRS/IN2P3, CEA/Irfu, Obs de Paris, Sorbonne Paris Cité, France
   \and
   Institute of Astronomy, University of Cambridge, Madingley Road CB3 0HA, Cambridge, UK 
   \and
   Institut d'Astrophysique de Paris, UMR 7095, CNRS, UPMC Univ. Paris 06, 98bis boulevard Arago, F-75014 Paris, France 
   \and
   Scuola Normale Superiore di Pisa - Piazza dei Cavalieri 7, 56126 Pisa, Italy 
   \and
   Università degli Studi di Urbino `Carlo Bo', Piazza della Repubblica 13, I-61029, Urbino, Italy 
   }

   \date{Received XXX; accepted YYY}

\authorrunning{Rossi et al.}
\titlerunning{The host galaxy of GRB 050219A}

 \abstract
{
 Long-duration gamma-ray bursts (LGRBs) are produced by the collapse of very massive
stars. Due to the short lifetime of their progenitors, LGRBs pinpoint star-forming
galaxies. Recent studies demonstrate that LGRBs populate all types of star-forming galaxies
from sub-luminous, blue compact dwarfs to luminous infrared galaxies.
}
{
We present here a multi-band search for  
the host galaxy of the long dark GRB 050219A within the enhanced \swift/XRT error circle.
We aim 
to characterise the properties of its host galaxy, and compare them 
with those of other LGRB host galaxies.
}
{
We used spectroscopic observations acquired with VLT/X-shooter 
to determine the redshift and star-formation rate of the 
 most probable host galaxy identified on the basis of a chance probability criterion.
We compared the results with the optical/IR spectral energy distribution obtained 
with \swift/UVOT, the seven-channel imager GROND at the 2.2-m telescope on La Silla
and the \textit{Herschel} Space Observatory, supplemented by archival observations obtained with FORS2 at the ESO/VLT, the \textit{Spitzer} Space Telescope and the \textit{GALEX} survey. 
}
{
The most probable host galaxy of the genuine long-duration GRB 050219A
is a 3 Gyr-old early-type galaxy at $z = 0.211$.
It is characterised by a 
ratio of
star-formation rate to stellar mass (specific star-formation rate)
of $\sim6\times10^{-12}\,{\rm yr}^{-1}$ 
that is unprecedentedly low when compared to all known LGRB host galaxies.
Its properties resemble those of post-starburst galaxies.
}
{
GRB 050219A might be the first known long burst 
to explode in a quiescent early-type galaxy.
This would be further evidence that GRBs can explode in all kinds of galaxies, 
with the only requirement being an episode of  high-mass star-formation. 
} 

\keywords{Gamma rays: bursts}

\maketitle

\section{Introduction\label{intro}}

The Gamma-Ray Burst (GRB) population is divided into two populations based on their duration: long GRBs
(LGRBs) and short GRBs \citep{Kouveliotou1993a}.
LGRBs are associated with the deaths of massive stars
mainly because of their association 
with broad-lined type Ic supernovae \citep[e.g.,][]{Hjorth2003a,Woosley2006a,HjorthBloom2012}.
Indeed, they can be found in all types of galaxies which feature star-forming activity (GRB host galaxies, GRBHs), 
from blue and low-mass \citep[e.g.,][]{Savaglio2009a} to red and massive galaxies 
\citep[e.g.,][]{Rossi2012a,Perley2013a,Hunt2014a}.

It has been a matter of debate whether GRBHs might be biased   
against particular properties such as dust content and metal abundance 
\citep[e.g.,][]{Elliott2012a,Michalowski2012a,Savaglio2012a,Perley2013a,Graham2013a,Kelly2014a}.
\citet{Jakobsson2012a} presented the redshift
distribution of the first optically unbiased LGRB host survey, which
included dark LGRBs\footnote{Here and in the following we consider dark GRBs to be those with an 
optical afterglow which is observed to be fainter than
what is predicted by the extrapolation from the X-rays \citep[e.g.,][]{Jakobsson2004a,VanderHorst2009a},
and following the standard afterglow theory \citep{Sari1998a}.} 
\citep{Hjorth2012a}. 
They showed that at $z\gtrsim3$ the LGRB rate does
not conform to the conventional determinations of the star-formation rate (SFR)
of the universe.
To investigate this,
\citet{Hunt2014a} combined the 
host samples compiled by \citet{Savaglio2009a} and \citet{Perley2013a}
and improved the sub-sample of the hosts of dark LGRBs by including 
new data and objects.
They find that LGRBs can explode in galaxies with a 
specific star-formation rate (sSFR$={\rm SFR}/M_*$) 
between $\sim10^{-10}$ and $\sim10^{-7}\,{\rm yr}^{-1}$.
These values are similar to the more general
 star-forming galaxy population up to $z\sim3$ \citep[e.g.,][]{Karim2011a}.

Here we present a comprehensive study of the putative host galaxy of the long GRB 050219A.
In particular we measure its mass and SFR and we show that
this galaxy has the lowest sSFR in comparison to the GRBHs known so far.
Throughout this work, 
we use a $\Lambda$CDM world model with $\rm \Omega_M$ = 0.27, $\rm
\Omega_\Lambda$ = 0.73, and $\rm H_0$ = 71 km$\,$s$^{-1}$Mpc$^{-1}$  \citep{Spergel2003}.
We use magnitudes in the AB system and the \citet{Chabrier2003a} 
initial mass function (IMF).
In the Appendix, we analyse the properties of the explosive event and afterglow
and show that GRB 050219A is truly a long burst.
We also review the evidence that GRB 050219A is a dark burst, and conclude
that it satisfies several criteria for such a classification.

\begin{figure}[tbp]
\begin{center}
\includegraphics[width=0.48\textwidth]{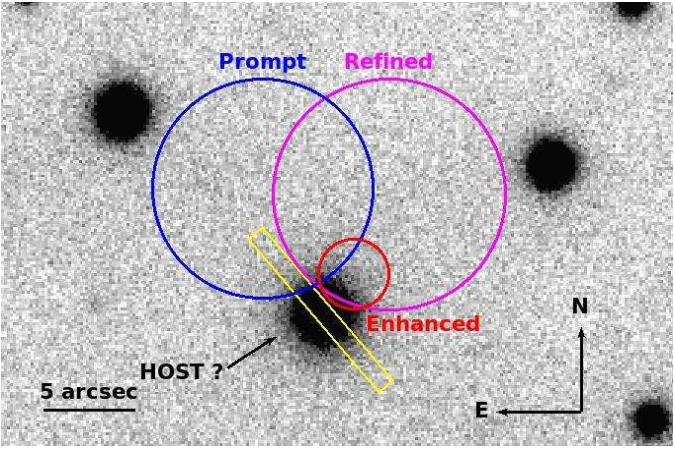}
\caption{ 
Zoom into the GROND $r^{\prime}$-band image of the field of GRB 050219A. 
We highlighted the prompt XRT position (blue), the refined position (magenta)
and the final enhanced position (red). The yellow rectangle shows 
the position of the X-shooter slit.
}
\label{fig:color}
\end{center}
\end{figure}

\section{Observations\label{obs}}

GRB 050219A triggered \swift/BAT at 12:40:01 UT on 19 February 2005 \citep{Hullinger2005GCN3038}.
It was a long burst with a duration of $T_{90}(15-350$ keV) $\sim24$ s.
\swift/XRT began observing 92 s after the trigger and found a bright, fading X-ray source \citep{Romano2005GCN3036}.
Simultaneous \swift/UVOT \citep{Roming2005a} 
observations started 80 s after the trigger and did not reveal any new source within
the prompt XRT error circle (see Fig.~\ref{fig:color}).
The first ground-based observations were performed $2.05$ hours after the event with the
Microlensing Observations in Astrophysics (MOA), an 0.6 m telescope at Mt. John University Observatory, New Zealand. 
The images did not reveal any variable source down to  $R = 20.5$ 
(\citealt{Postigo2005GCN3041}; see also \citealt{Berger2005GCN3048}).

Two years later a new enhanced XRT position \citep{Goad2007a,Evans2009a}
was published with a localisation error of $1\farcs9$. 
Afterwards, the position was revised but remained consistent within
$0\farcs5$. The latest position available as of July 2014 on the {\it UK Swift Science Data Centre}
\footnote{ See {\tt www.swift.ac.uk/xrt\_positions}.
The webpage offers a brief introduction \citep[but see also][]{Evans2009a}.}
is centred at coordinates RA, Dec (J2000) = $11^{\rm h}05^{\rm m}38\fs 97$, $-40^\circ41'02\farcs6$
 with an error of $1\farcs9$, and about $6''$ offset from the initial XRT position (Fig.~\ref{fig:color}).
Our analysis of the XRT emission compared with the optical upper limit reveals that
this was a dark GRB (see Sect.~\ref{app:dark}), 
therefore we included it in our survey aimed at searching 
for the hosts of this class of bursts \citep{Rossi2012a}.
Afterwards, we imaged the field with the Gamma-Ray burst Optical \& Near-infrared
Detector \citep[GROND;][]{Greiner2008a}, a seven-channel imager mounted on the 2.2m MPG/ESO
telescope on La Silla (Chile).
We found a bright extended source partly covered by the enhanced error circle.
We re-examined the MOA $R$-band images and found no source at the new position except the 
extended source (which is just barely visible), 
 but it is possible that the large seeing ($4''-5''$) prevents us from
distinguishing the afterglow from this object.
This galaxy is also visible in the Palomar Digitized Sky Survey. 
We identify it as a host galaxy candidate of GRB 050219A (GRBH 050219A; see Fig.~\ref{fig:color}).

To obtain a more complete view of the spectral energy distribution (SED)
of the host candidate we searched the UVOT images obtained during the day of the GRB trigger.
The UVOT $v$-band host detection is similar in depth to the MOA upper limit,
thus we consider it unlikely that the afterglow could dominate the host in the UVOT $v$-band.
Additionally, we retrieved and analyzed archival FORS2/$R$-band images
(ESO/VLT) and {\it Spitzer Space Telescope} observations. 
The field was also covered by the {\it Galaxy Evolution Explorer} (GALEX) UV survey
\footnote{The field was covered by the {\it Wide-Field Infrared Survey Explorer} 
(WISE) and the galaxy is detected in the W1-band ($3.4\, \mu m$) but with 
a lower significance than in the IRAC $3.6\, \mu m$ band. 
Therefore, we do not consider this detection here.}. 
Furthermore, members of our team were awarded observing time to target the host candidate with the {\it Herschel Space Observatory} \citep[][]{Hunt2014a}.
The summary of the observations is given in Table~\ref{tab:phot}.

Finally, on the 19th of April 2012 we obtained UV to NIR spectroscopic observations of the host galaxy candidate
with the X-shooter instrument \citep{Vernet2011a} mounted on the VLT on Paranal (ESO, 
Chile)\footnote{The observations were conducted as a part of the 
Italian Guaranteed Time under program 089.A-0843(A) (PI: S. Piranomonte).}. 
The observations were performed at seeing $\sim1$\arcsec\ 
and airmass $\sim1.1$. The slit was placed along the parallactic angle
which was at $\sim40$ degrees counterclockwise from the north, as shown in Fig. \ref{fig:color}.
They
consisted of four exposures with an exposure time of 600 s each. They were obtained by nodding along the 
slit with an offset of $5''$ between exposures in a standard ABBA sequence. We used slit widths  of 
$1\farcs0$, $0\farcs9$ and $0\farcs9$ for the UVB, VIS and NIR spectrograph arms, respectively, 
resulting in resolving powers of $R = \lambda / \Delta \lambda \approx 4400, 7400$ and $5400$.

\section{Data analysis\label{sec:dataana}}

\swift/UVOT data were analysed using the standard analysis
software distributed within FTOOLS, version 6.5.1. The source count rates 
were extracted within a $3''$ aperture. 
An aperture correction was estimated from selected
nearby point sources in each exposure and applied to obtain the
standard UVOT photometry calibrated for a $5''$ aperture.

GROND $g^{\prime} r^{\prime} i^{\prime} z^{\prime}$ and $JHK_s$ images were reduced in a
standard manner using PyRAF/IRAF \citep{Tody1993}.
The procedure is based on the pipeline written
to reduce GROND data \citep{KupcuYoldas2008AIPC,Kruhler2008a}. 
Photometry was performed
by using an aperture diameter of $7$\arcsec, that is $2.5$ times the full width at half
maximum (FWHM) of the stellar PSF in $K$-band (the largest value among the GROND filters). 
Within this aperture the measured flux 
flattened in a curve-of-growth analysis and therefore it 
is sufficiently large to include all of the
galaxy flux.  
GROND optical bands were calibrated against the secondary photometric standards 
listed in Table \ref{tab:std}. Their magnitudes were derived from observations
of SDSS fields. GROND NIR bands were calibrated against 2MASS field stars.
The FORS2/$R$-band image was analysed in the same way but calibrated using 
a zeropoint given by ESO.
In the following, we will not use the FORS2
photometry because the accuracy of the GROND photometric calibration is better
than the one available for the FORS2/$R$ band. 
UV, optical and NIR magnitudes were corrected for Galactic
extinction using the interstellar extinction curve derived by
\citet{Cardelli1989} and by assuming $E(B-V)=0.16$ mag \citep{Schlegel1998}
and a visual extinction to reddening ratio of $R_V=3.1$.
The photometry on {\it Spitzer} and {\it Herschel} 
data were performed with an aperture larger than $9$ arcsec. 
The method is described by \citet{Hunt2014a}.

The host galaxy candidate is detected in the UVOT $v$-band, all GROND optical/NIR bands, the FORS2/$R$-band,
and with both {\it Spitzer} IRAC and MIPS.
The galaxy is not detected in the UV (GALEX and UVOT) or in 
{\it Herschel} bands. 
The summary is given in Table~\ref{tab:phot}.
The coordinates of the host galaxy candidate are RA, Dec (J2000) = 
$11^{\rm h}05^{\rm m}39\fs 07$, $-40^\circ41'04\farcs6$.
They are derived from the GROND $r'$-band which have an 
astrometric precision of about $0\farcs3$ corresponding to the rms accuracy of the USNO-B1 catalogue \citep{Monet2003}.
Given coordinates of the new enhanced XRT position (see above)
are offset $\sim2''$ from the galaxy centre.

We processed the X-shooter spectra using version 2.0.0 of the data
reduction pipeline \citep{Goldoni2006a,Modigliani2010a}, 
using the reduction technique developed for nodded observations. 
To flux-calibrate the spectrum we used the observations of the spectrophotometric standard star LTT 3218, taken 
in the nodding mode.

Flux values were corrected for Galactic extinction.
Afterwards, we cross-checked and calibrated the fluxes
using the corresponding magnitudes 
of the galaxy (see Tab.~\ref{tab:phot})
and therefore accounting for slit-aperture flux losses, 
because the slit was much smaller than the entire galaxy.
The final correction factor is $6.2$.


\section{Results \label{sec:prop}}

\begin{figure}[tbp]
\begin{center}
\includegraphics[width=0.48\textwidth]{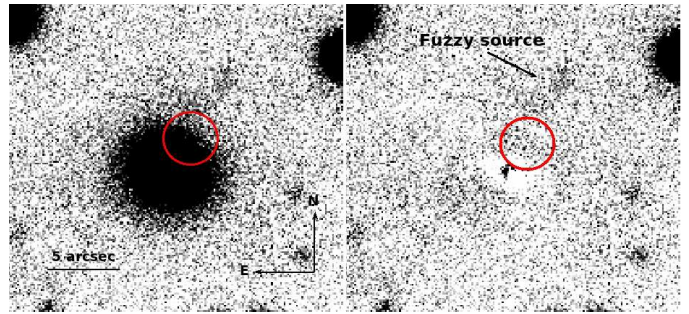}
\caption{ 
{\bf  Left:} Zoom into the FORS2/$R$-band image centred on the enhanced XRT error circle (in red) of GRB 050219A. 
{\bf  Right:} The subtraction of the bright galaxy with {\tt GALFIT} 
does not reveal any source within the XRT error circle. However, a fuzzy source is visible north-west and outside 
the XRT error circle (see Sect.~\ref{sec:imasub} for details). The bright spot at the center of the image is just a 
residual of the template subtraction (south-eastern border of the XRT circle).
}
\label{fig:imasub}
\end{center}
\end{figure}

\begin{figure*}[!tbp]
\begin{center}
\includegraphics[angle=-0,width=0.99\textwidth]{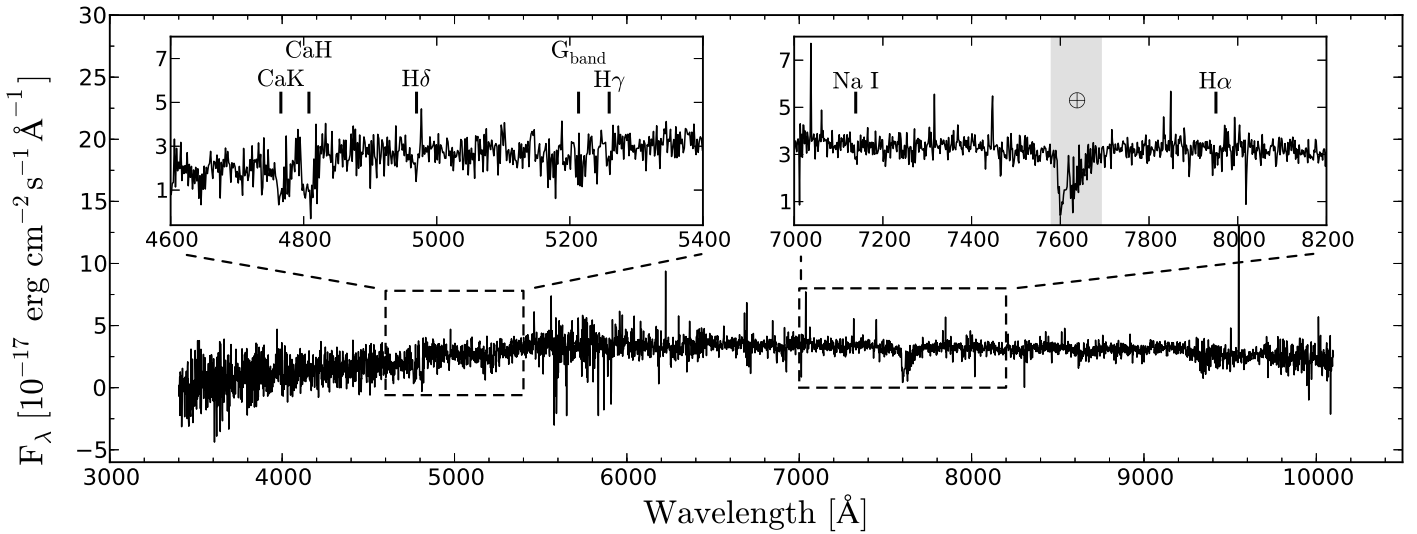}
\caption{ X-shooter (UVB and VIS arms) spectrum of the putative host galaxy of GRB 050219A (Fig.~\ref{fig:color}).
Several absorption lines are detected at a common redshift of $z=0.211$.
The zoom-in panels show the regions where the absorption features are visible. 
Only continuum is detected in the infrared X-shooter arm, therefore it is not presented in the plot.
}
\label{fig:xshooter}
\end{center}
\end{figure*}


\subsection{The host galaxy candidate and its morphology \label{sec:imasub}} 

The host candidate has a featureless morphology 
with a FWHM of $\sim1\farcs7$ both in the GROND/$r^{\prime}$ and the FORS2/$R$-band images, 
i.e.,  larger than the stellar FWHM ($\sim1$\arcsec\ for both observations), 
thus it is clearly a galaxy. 
The galaxy has a small elongation along the NE-SW axis ($a/b\sim1.1$, 
where $a$ and $b$ are the semi-major and the semi-minor axis of the best matching ellipse)
and an averaged half-light radius of $\sim1\farcs3$.

 No other close-by galaxies are visible in any of the images, 
 and the few other objects within a few arcsec are stars or they are too faint
to be clearly classified as galaxies. 
One of them is a fuzzy source barely visible in the FORS2/$R$-band image 
outside the north-west border of the XRT error circle and offset $4\farcs6$ from its center (Fig.~\ref{fig:imasub}).
The source is detected at only $2\sigma$ confidence level, thus it might not be real.
If real it could be a galaxy, perhaps a 
small companion of the bright host candidate. 
Aperture photometry for this fuzzy source gives an extinction-corrected magnitude of 
$R\sim25.5\pm0.6$, beyond the FORS2 $3\sigma$ limiting-magnitude of $\sim25.0$.

If there were another galaxy within the XRT error circle, 
the glare of the bright galaxy makes its identification very hard.
Therefore, we used {\tt GALFIT v.3.0.5} \citep{Peng2010a} to subtract the 
bright galaxy (within the XRT error circle) in the FORS2/$R$-band image 
and search for possible hidden objects. We obtained the best solution using 
a 2D S\'ersic profile with an index $n=3.3\pm0.2$ and an effective radius 
equal to its half-light radius ($1\farcs3$; see above).
The profiles were convolved with the stellar PSF. 
To build the PSF we measured the magnitudes of isolated stars
with ten different apertures with a radius in the interval (FWHM, $2\times$FWHM). 
Afterwards, we fitted their profile
using the {\tt psf} routine under the {\tt DAOPHOT}/IRAF package,
following the prescriptions outlined in \citet{Stetson1987a}
and in the {\tt DAOPHOT} manual. The PSF image model to be used in {\tt GALFIT} 
was obtained with the task {\tt seepsf} and finally normalized.
After subtraction, no sources are detected within the enhanced XRT error circle (right panel in Fig.~\ref{fig:imasub}). 

Because of the featureless morphology and the best fit with a S\'ersic profile
having $n>2$, 
we classify the host candidate as an early-type galaxy \citep[ETG; e.g.,][]{Glazebrook1995a,Rowlands2012a}.

\begin{figure}[!tbp]
\begin{center}
\includegraphics[angle=-0,width=0.45\textwidth]{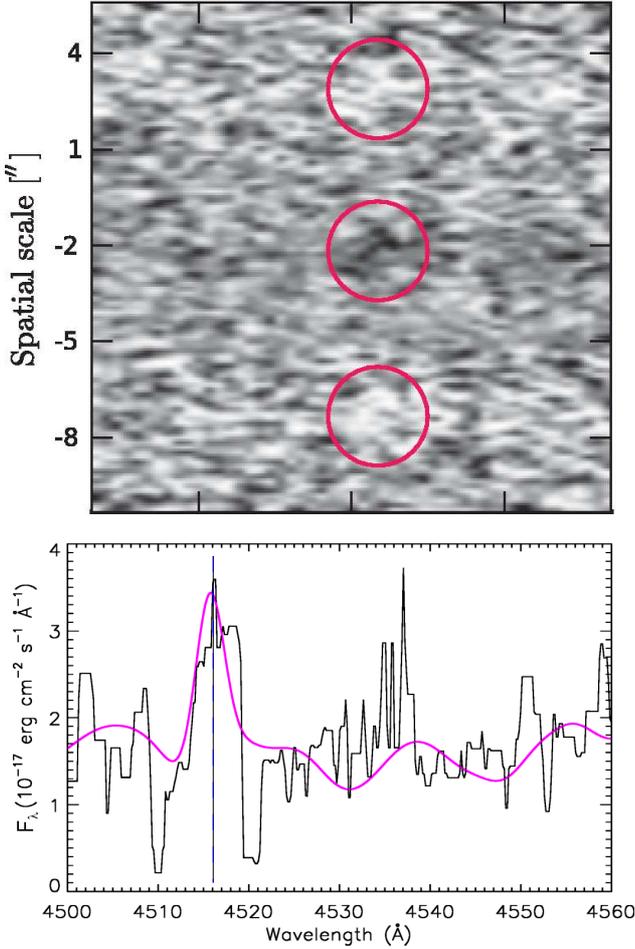}
\caption{ 
Zoom-in of the region occupied by the [\ion{O}{ii}] emission line  in the X-shooter spectrum. 
{\textbf{{Top}}:} 2D spectrum. The circles mark the emission line and the negative features resulting from the nodding. 
{\textbf{{Bottom}}:} 
Modeling of the spectrum obtained with {\tt GANDALF} (magenta line) compared to
the one-dimensional spectrum. The spectrum has been median filtered, flux-calibrated, 
and corrected for Galactic extinction.
The [\ion{O}{ii}] emission line is redshifted to $4516$ {\AA} (blue dashed line) 
in agreement with the redshift of $0.211$ measured with the absorption lines only. 
}
\label{fig:OII}
\end{center}
\end{figure}


\subsection{Spectroscopic redshift and star formation \label{sec:spec}}

In the X-shooter spectrum, the galaxy continuum is well visible and there are clear detections of 
absorption lines typical of an elliptical galaxy with an old stellar population (Ca H\&K, G-band, and 
several Balmer absorption lines including H$\alpha$, H$\delta$, and H$\gamma$) 
at redshift $z=0.211$ (see Fig.~\ref{fig:xshooter}).
At this redshift, 1 arcsec on the sky corresponds to $3.4$ kpc, the distance modulus is $m - M = 40.07$ mag,
and the look-back time is 2.57 Gyr (11.15 Gyr after the Big Bang).
The half-light radius of the galaxy ($\sim1\farcs3$) corresponds to $\sim4$ kpc
and its projected offset from the centre of the XRT position is $\sim$6 kpc.
We find a 3$\sigma$ flux excess 
at the position of the
[\ion{O}{ii}] emission line (Fig.~\ref{fig:OII}). 
The spectrum does not allow us to resolve the [\ion{O}{ii}] doublet.
This is due to the poor signal-to-noise ratio and 
the dispersion velocity of the gas in the galaxy
that is comparable to the relative separation of the two 
lines of the [\ion{O}{ii}] doublet. 
We find a flux (calibrated and corrected for the aperture; Sect.~\ref{sec:dataana})
of $8.7 \times 10^{-17}$ erg cm$^{-2}$ s$^{-1}$, which corresponds to a
star-formation rate (SFR) of $0.06_{-0.02}^{+0.01} \, {\it M}_{\odot}\,{\rm yr}^{-1}$
 following the prescription given in \citet{Savaglio2009a}.
  Note that the slit aperture does not intersect 
the center of the XRT error circle, where there could be additional 
star formation, signalled by the explosion of the long GRB. 
However, this star formation should be visible in the SED.

\subsection{Modelling of the spectrum \label{sec:gandalf}}

We modelled the X-shooter spectrum of the host candidate using 
 the {\tt GANDALF} software \citep{Sarzi2006a} to
investigate the likelihood of the detection of the [\ion{O}{ii}] line.
{\tt GANDALF} linearly combines
a set of stellar templates convolved with a line-of-sight velocity
dispersion and fits galaxy spectra \citep{Capp2004a}.
At the same time it also fits a user-defined list of gas emission
lines modelled with Gaussian templates. As stellar templates we used
a subset of the X-shooter Spectral Library \citep[XSL;][]{ChenYP2014a}.
The input spectrum was flux-calibrated and de-reddened as described above. 
It was then rebinned logarithmically to have a constant velocity dispersion
and fit between 3600 and 4100 {\AA} in the rest frame. This spectral
region was chosen because of the presence of [\ion{O}{ii}] and of the
strong Ca II H \& K feature which helps in the convergence of the fit.
We obtained a best fit stellar continuum with a combination of
G, K and M stars and the [\ion{O}{ii}] line  was found in the best fit 
model with a flux of  7.1 $\pm$ 2.2 $\times 10^{-17}$ erg cm$^{-2}$ s$^{-1}$ 
(Fig.~\ref{fig:OII}, bottom panel). No further extinction is necessary to fit the spectrum. 
We thus confirm the presence of the [\ion{O}{ii}] line at the $\sim$3$\sigma$ level.

H$\alpha$  and H$\beta$ emissions are not detected in our spectrum.
Balmer lines are 
more uncertain than [\ion{O}{ii}] 
due to the undetermined underlying stellar absorption, which is more important 
 in older stellar populations.
 The H$\alpha$, H$\beta$, and [\ion{O}{iii}] emission lines are constrained by 
 {\tt GANDALF} to  $< 7 \times 10^{-17}$ erg cm$^{-2}$ s$^{-1}$ at $3\sigma$, 
therefore confirming their non-detection in the X-shooter spectrum.
The upper limit on H$\alpha$ corresponds to a SFR $\lesssim0.04 \, {\it M}_{\odot}\,{\rm yr}^{-1}$,
consistent with the SFR determined with [OII]. The other upper limits are not constraining. 
However, we prefer to take the given upper limits with caution because 
H$\alpha$ and H$\beta$ emission lines are in a noisy part of the spectrum and fall within the corresponding absorbing lines,
 decreasing the chance of detection.
Therefore, given our data set, the best SFR estimator is the [\ion{O}{ii}] emission-line.

The non-detection of the Balmer emission lines prevents us from constraining the extinction; 
however our analysis of the spectrum with {\tt GANDALF} showed no
evidence for extinction by dust.
This is not surprising, because there are intriguing examples
of dust-extinguished LGRBs hosted in blue and young galaxies with very low global extinction \citep[e.g.,][]{Kruhler2011a}.
In general, the global dust extinction in GRBHs is in many cases negligible \citep[e.g.,][]{Hunt2014a}.
This is further supported by the low average extinction measured along GRB sight-lines ($A_V<0.4$ mag) and the fact that
afterglows with negligible $A_V$ are common \citep[e.g.,][]{Kann2010a,Covino2013a}. 
Taken together, these cases illustrate that the star-forming regions 
typically associated with LGRBs have patchy dust distributions.
Therefore, the dust extinction depends on the geometry of the dust distribution, 
and not on the global properties of the host galaxy. 
However, note that even if we assume a reasonable $A_V = 1$~mag in the star-forming region 
in agreement with the dark nature of GRB 050219A, the SFR would be just a factor of $2.5$ higher.

\begin{figure}[tbp!]
\centering{
\includegraphics[width=0.48\textwidth,angle=0]{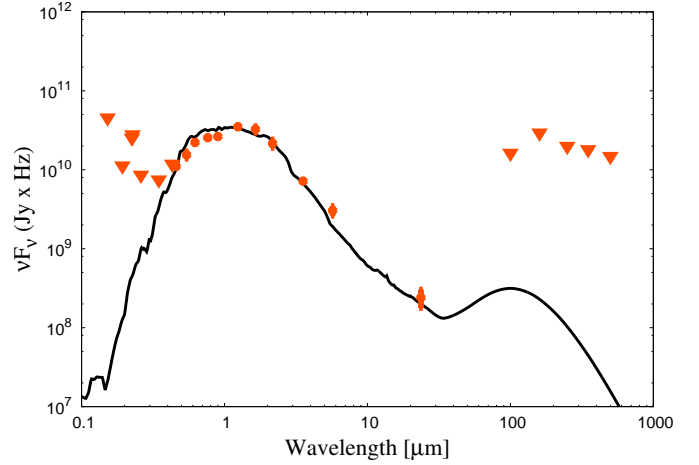}
}
\caption{
 SED fitting of the host galaxy candidate of GRB 050219A with {\tt GraSil} from UV ({\it GALEX}) to sub-mm wavelengths ({\it Herschel}).
The SED is best fit by a galaxy template with the following properties: age $\sim3\,{\rm Gyr}$, 
SFR $\lesssim0.1\, {\it M}_{\odot}\,{\rm yr}^{-1}$, and stellar mass M$_*\sim10^{9.98}\,{\it M}_\odot$. 
The IR star-formation and dust mass are not constrained by the observed 
{\it Herschel} upper limits. 
}
\label{fig:sed}
\end{figure}


\subsection{Modelling of the spectral energy distribution \label{sec:sed}}

The spectral energy distribution of the host candidate of GRB 050219A 
has been modelled making use of template SEDs developed with
{\tt GraSil}\footnote{\tt www.adlibitum.oat.ts.astro.it/silva} \citep{Silva1998a} 
and using the SED-fitting procedure described in \citet{LoFaro2013a}.
Differently from semi-empirical approaches {\tt GraSil} is a self-consistent 
physical model which allows to predict the SEDs of galaxies from the far-UV
to radio including a state-of-the-art treatment of dust extinction 
and reprocessing based on a full radiative transfer solution.
Moreover, it includes star-formation histories (SFHs) which are 
self-consistently computed following the chemical evolution of the galaxy. 
The photometric data of the host candidate
are best-fit by a galaxy with a $3\,{\rm Gyr}$-old stellar population with 
a stellar mass of $10^{9.98}\,{\it M}_\odot$ and negligible dust content $A_V<0.1$ mag (Fig.~\ref{fig:sed}). 
We also find the photometric data set to be consistent with a SFR $\lesssim\,0.1\,{\it M}_{\odot}\,{\rm yr}^{-1}$.
These results are in good agreement with a previous SED fit presented by \citet{Hunt2014a},
who used an older set of SED templates and finds a stellar mass of $10^{9.91}\,{\it M}_\odot$. 
We conclude that the measurement of the [\ion{O}{ii}] line provides a good estimate of the current SFR.
We noted in section~\ref{sec:spec} that
the X-shooter slit does not intersect 
the center of the XRT circle, and there could be additional 
star formation. However, our aperture photometry covers the whole galaxy and
partly the XRT circle, and any star-forming region in the galaxy 
contributes to the observed SED, including those not covered by the X-shooter slit. 
Therefore, the SED fitting provides an upper limit to the global SFR of the host candidate.

The parameters estimated via SED-fitting are notoriously affected by degeneracy 
and it is therefore difficult to estimate their errors. In particular, age and 
stellar masses tend to differ according to the adopted SFHs. The best fit template is based on a SFH
with an infall time-scale $\tau=0.1$ Gyr and an efficiency of 2.3, 
typical of galaxies dominated by an old stellar population \citep[e.g.,][]{Silva1998a}. With this SFH, we have obtained a 
reasonable result in the age range of $3_{-0.5}^{+1}$ Gyr. 
A discussion of the derivation of galaxy properties and typical uncertainties
is given by \citet{Michalowski2010a,Michalowski2012b,Michalowski2014a}.

\subsection{Is there AGN activity? \label{sec:agn}}  

Since the [\ion{O}{ii}] emission line is also observed in AGNs, 
one might wonder whether the line observed in the galaxy spectrum
is due to nuclear activity.
Former studies have shown that
GRBHs do not generally show
signatures of AGN activity in 
their emission lines \citep{Watson2011a} and 
in their SEDs \citep{Michalowski2008a}.
Note that [\ion{O}{ii}] alone is not sufficient to claim the presence of nuclear activity.
but other lines are necessary. 
In particular, there is no  [\ion{O}{iii}] emission visible in the spectrum, which 
in AGNs is usually stronger or at least 
comparable to the flux observed in [\ion{O}{ii}] \citep[e.g.,][]{Kewley2007a}.
X-ray emission is one of the principal characteristics
of AGN activity. No X-ray source is known at the position of the bright host candidate or in the 10 arcmin surrounding radius.
The \swift/XRT observations provide an upper limit of $\sim1.7\times10^{-19}$ erg cm$^{-2}$ s$^{-1}$ 
(see the \swift/XRT on-line repository\footnote{{\tt http://www.swift.ac.uk/xrt\_spectra/}; 
\cite{Evans2007a,Evans2009a}.}).
At $z=0.211$ this corresponds to a luminosity ${\rm L}_X < 2.4\times10^{42}$ erg s$^{-1}$, 
which is enough to constrain the presence of an AGN. 
However, note that AGN activity affects the mid-IR SED with a power-law component that
 is not detected in the observed SED (see Sect.~\ref{sec:sed}).
Therefore, the [\ion{O}{ii}] emission line 
is most likely due to on-going star-formation rather than nuclear activity.


\section{Discussion \label{disc}}

\subsection{Comparison with apparently similar cases \label{sec:others}} 

If GRB 050219A were a short GRB \citep[e.g.,][]{Berger2013b}, 
the association with a galaxy having a low SFR
dominated by an old stellar population would not be unexpected; 
more than $50\%$ of all short GRBHs have an evolved stellar 
population, and ETGs are common \citep[e.g.,][]{Leibler2010a,Berger2011a}.
However, the possibility that GRB 050219A is a short burst is excluded,
 because not only does it have a long duration ($T_{90}\sim24$~s) 
but it also satisfies the ${\rm E}_{\rm p,i} - {\rm E}_{\rm iso}$ and the lag-luminosity correlations 
for LGRBs (Sect.~\ref{app:prompt}).

The case studied here is apparently similar to the LGRB 060912A 
\citep{Levan2007a} and the LGRB 130702A \citep{Kelly2013a}. 
In the first case, the
initial images showed that the galaxy closest
 to the GRB position was a low-redshift elliptical galaxy $10''$ away from the nominal locus. 
 The situation changed when the deepest VLT imaging was acquired, showing that the real host is a 
 star-forming galaxy at $z=0.937$ lying within the XRT error circle. 
In the case of the LGRB 130702A, the explosion site is offset $\sim7\farcs6$ from the center
of a bright red disk-dominated galaxy, but $\sim0\farcs6$ away from the center of a much fainter 
metal-poor dwarf galaxy. Both galaxies have the same spectroscopic redshift as the GRB afterglow ($z=0.145$) 
and form a gravitationally bound system. \citet{Kelly2013a} concluded 
that the dwarf galaxy is the most likely host of LGRB 130702A on the basis of the smaller offset
from the GRB site. 

The case of GRB 050219A represents the converse situation: the bright ETG host candidate
is the closest (in projection) to the XRT error circle and it partially 
overlaps with the circle.
Therefore, the chance probability of a misidentification is much lower.

\citet{Kelly2013a}  argued that GRB 130702A is a lucky case where the red massive galaxy and 
its dwarf satellite (i.e., the GRB host) are well separated in the sky,
but the same situation can result, in some cases, in a superposition along the sight line
and in this case it would be a lot harder to distinguish the two galaxies.
We have shown that no other candidates are detected even after image subtraction.
In the following we will investigate the properties that 
a candidate host should have in order to not be detected 
in the case of GRB 050219A.
If we take the dwarf host of GRB 130702A as a typical dwarf GRBH at low redshift 
we can calculate how it would appear at $z=0.211$. 
In order 
to keep the same absolute magnitude ($M_r\sim-16$~mag),
it would have an apparent magnitude $r\sim24.0$~mag. Therefore, it would be 
clearly detected
in the FORS2 image that has a $3\sigma$ limiting magnitude of $R\sim25$~mag. 
To have the same $R$-band magnitude as the fuzzy object, the dwarf GRBH 130702A would be at 
$z\sim0.45$.
 Of course a dwarf galaxy, like the putative companion of the bright host candidate 
 can even be fainter than GRBH 130702A.
We conclude that an alternative host candidate for 050219A 
should be redder and/or fainter than GRBH 130702A,
or should lie at a redshift significantly higher than $z=0.211$.


\begin{figure*}[!h!t]
\centerline{
\vspace{0.5cm}
\includegraphics[angle=0,height=0.46\linewidth,bb=18 160 592 650]{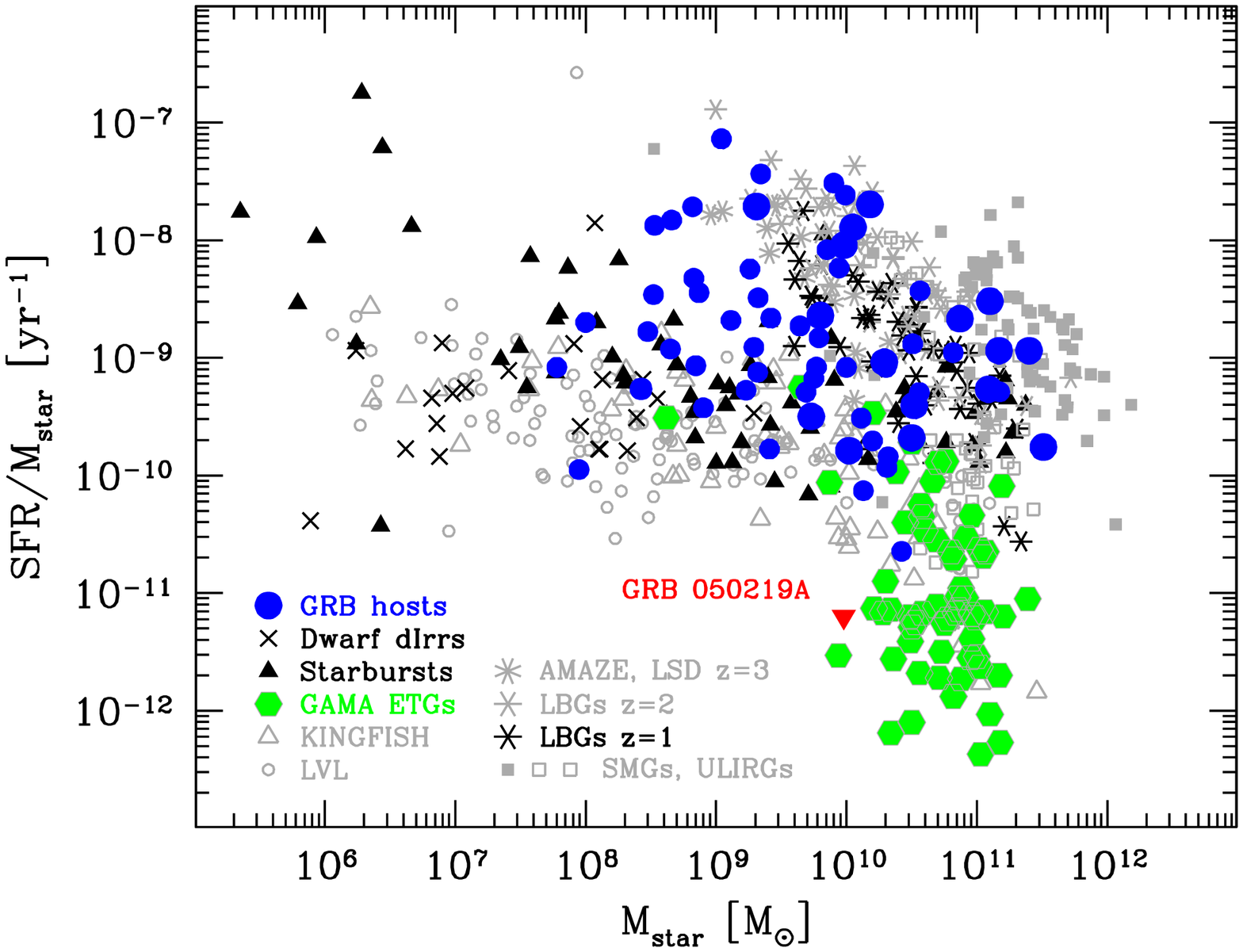}
\hspace{-1.9cm}
\includegraphics[angle=0,height=0.46\linewidth,bb=18 160 592 650]{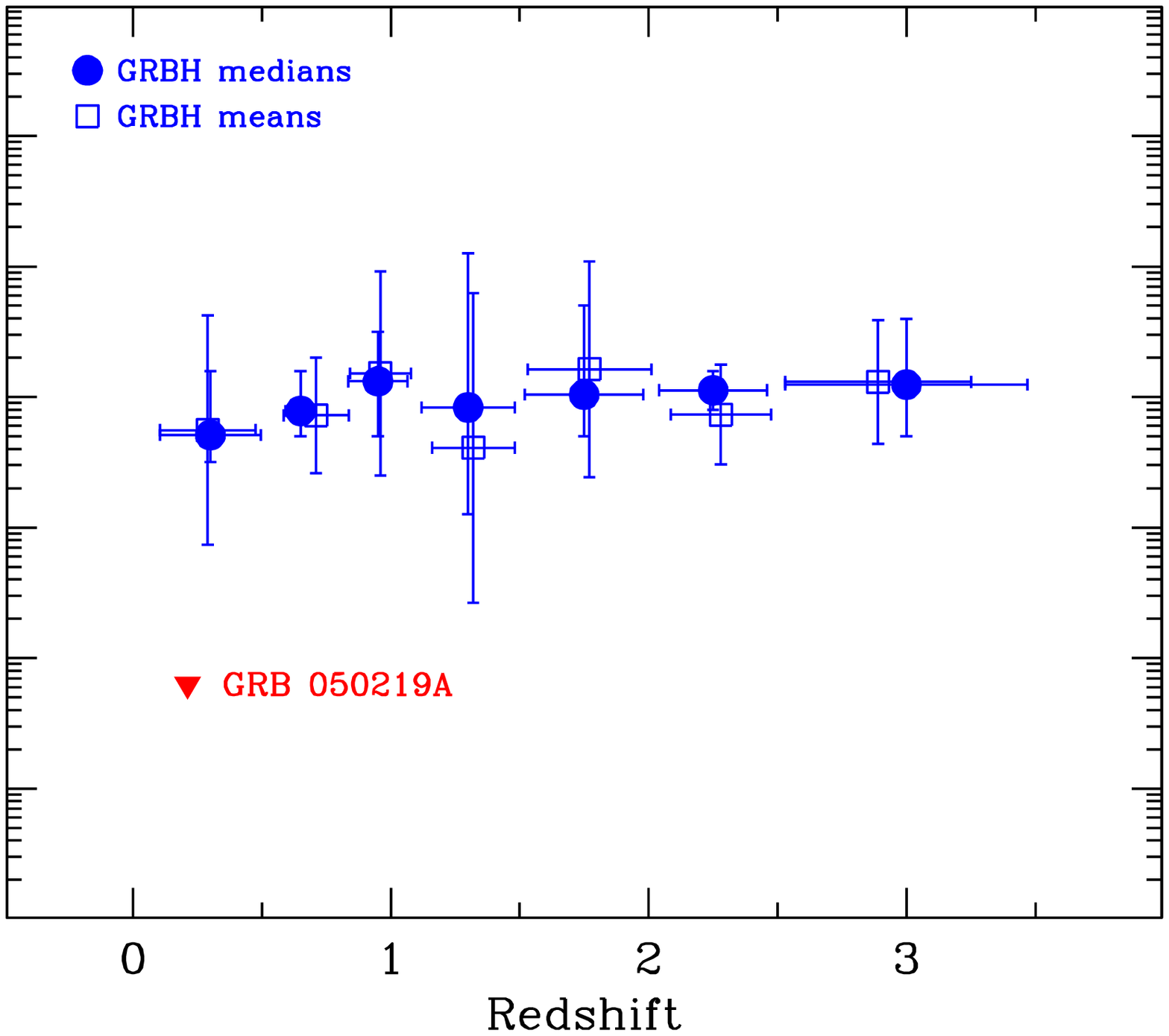}
}
\caption{
The host galaxy of GRB 050219A compared to other GRBHs and other populations of galaxies
\citep[adapted from][]{Hunt2014a}.
\textbf{{Left}}:
 sSFR vs. stellar mass. The
host of GRB 050219A (marked in red) lies in a region occupied by ETGs from the GAMA survey 
\citep[][marked in green]{Rowlands2012a}. The best-studied GRBHs are marked in blue.
\textbf{{Right}}:
sSFR plotted against redshift 
for the GRBH sample from Hunt et al. (2014).
The filled circles show the medians of each redshift bin,
and the vertical error bars correspond to the
upper and lower quartiles of the GRBH distributions.
The empty squares show the mean of the 
GRBH distributions within each redshift bin, and the error bars correspond to the
standard deviation.
The horizontal error bars are the width of the
redshift bins (for the GRBH data).
\label{fig:sfr}
}
\end{figure*} 


\subsection{How likely is the association of the host candidate with GRB 050219A? \label{sec:prob}}

GRB 050219A was selected from a parent sample of optically faint GRBs 
but with \swift/XRT detections. 
The details of selection criteria are explained in \citet{Rossi2012a}. 
About 40 targets fulfil these selection criteria.
However, due to the limited amount of observing time the sample
was restricted to 18 targets. 
We caution that the sample 
is not complete because it is biased towards dark GRBs
and not representative of the whole GRBH population. It is therefore difficult
to estimate the probability of a random association of dark GRBs with a particular class 
of galaxies like ETGs.

In general, the only way to unambiguously associate a host candidate with a GRB is to obtain
the redshifts of afterglow and galaxy and find them to be identical. 
When no optical afterglow is seen, as is typical for dark GRBs, 
the host is chosen on the basis of positional coincidence and the 
chance probability \citep{Bloom2002a}. It is also possible to do this by relying on the 
properties of the candidate host(s), including colour and SFR, 
in order to identify star-forming galaxies \citep[e.g.,][]{Rossi2012a}.

\citet{Bloom2002a} introduced a criterion to calculate the
probability $p$ of finding a galaxy of given 
(extinction-corrected) $R$-band magnitude in a circular region of radius $r$.
These authors discuss three possible scenarios for determining the radius: i) the GRB is well localized 
inside the detectable light of a galaxy; ii) the localisation is poor
and the host candidate is within the error circle; iii) the localisation is good but it 
is outside the light of the nearest galaxy. 
The most appropriate scenario is the one that gives the largest radius.
The case of the putative host of GRB 050219A is in between 
scenarios ii) and iii) and the radius is $r=3\farcs5$, given by the $3\sigma$ confidence level 
of the XRT localisation error. 
Following \citet{Bloom2002a}, we used the $R$-band counts of galaxies brighter than $27$ mag of a field 
at high galactic latitude presented by \citet{Hogg1997}. 
We find that the chance probability ($p$-value) that a galaxy with an equivalent (or
brighter) magnitude ($R\sim20$) lies within a $3\farcs5$ radius of the XRT position
is $\sim0.8\%$.
For the fuzzy source possibly detected in the FORS2 image with $R\sim25.5$, the radius is given by 
the third scenario described above. Following \citet{Bloom2002a}, the radius is given by 
$\sqrt{r_{of}^2+r_{half}^2}=5''$, where $r_{of}$ is the offset of the centre of the source 
from the XRT position ($4\farcs6$) and $r_{half}$ is the half-light radius of the source ($\sim1''$).
 For the fuzzy source we find a $p$-value of $\sim68\%$.

We conclude that the small chance probability for the bright candidate 
and the high probability for the fuzzy object supports the idea that
the bright galaxy in the XRT error circle is the best host candidate.
and in the following we will simply refer to it as the {\it host galaxy candidate}, or GRBH 050219A.


\begin{table*}[!htbp]
\begin{center}
\caption{Summary of the spectral properties of GRBH 050219A and other post-starburst galaxies}
\begin{tabular}{lccccccc}
\toprule
          &          & EW(H$\delta$) & EW[\ion{O}{ii}] &                    &  $g'-r'$      & SFR$([\ion{O}{ii}])$ & $M_*/$SFR \\
Galaxy    & redshift & ({\AA})         & ({\AA})           & log($M_*/M_\odot$) & mag & ($M_\odot\,{\rm yr}^{-1}$) & (Gyr) \\
\midrule
GRBH 050219A  & 0.211 & 2.01   & $-$3.17  & 9.98   & 0.93 &	$0.06$   & 170   \\
GDDS-12-8139  & 1.189  &8.36   & $-$6.21   & 10.39  & 0.40 & 2.3      & 10.5  \\       
GDDS-02-0715  & 1.133  &3.01   & $-$4.29  & 11.25  & 0.37 &   1.8     &101.6  \\
GDDS-02-1543  & 1.131  &4.22   & $-$5.34   & 10.80  & 0.67 &0.2        &267.4  \\
GDDS-12-8983  & 0.963  &0.72   & $-$1.51   & 10.70  & 0.56 &0.6        &89.7   \\
\bottomrule	 
\end{tabular}
\tablefoot{The post-starburst  galaxies are taken from \cite{LeBorgne2006a}.}    
\label{tab:spec}
\end{center}
\end{table*}

\subsection{The host galaxy candidate of GRB 050219A compared with general galaxy populations}  

It is justified to assume that the progenitor of the LGRB 050219A was a massive star which
exploded as a broad-line type Ib/c supernovae 
\citep[SN Ic-BL, associated with LGRBs, e.g.,][]{Bersier2012a,Schulze2014a}. 
Recently, \citet{Kelly2014a} compared hosts of LGRBs and SNe Ic-BL
to those of core-collapse SNe from the SDSS survey \citep{Ahn2014a}. 
Compared to \citet[][their figure 10]{Kelly2014a} GRBH 050219A has a lower sSFR,
and with a half-light radius of $\sim4$ kpc (see Sect.~\ref{sec:prop}) it is larger than GRBHs and galaxies hosting SNe Ic-BL which 
usually have a radius $r\lesssim2$ kpc \citep[][their Fig. 8]{Kelly2014a}.
However, the host of SN 2010ah has a similarly low SFR ($\rm{SFR}\sim0.1\,{\it M}_\odot\,{\rm yr}^{-1}$) 
and radius ($\sim4.4$ kpc). 

\citet{Hunt2014a} present new {\it Herschel}/far-IR data for 17 host galaxies,
where 14 of them are hosts of dark GRBs. 
These galaxies are then included in a larger sample of 66 hosts
combining those presented by \citet{Savaglio2009a} and \citet{Perley2013a}.
In Figure~\ref{fig:sfr} we show the sSFR vs. stellar mass and redshift from \citet{Hunt2014a} 
and highlight GRBH 050219A.
 In particular the right panel shows mean and median of the sSFR for each redshift bin 
of the GRBH distribution, which are consistent within the errors. 
GRBH 050219A stands out as an isolated case because it has the lowest 
specific SFR among all the known hosts to date 
($\rm{sSFR}\lesssim10^{-11}\,{\rm yr}^{-1}$, using the result from the SED fitting).
This low sSFR corresponds to UV brightness at $0.2\mu m$ of $\gtrsim24$ mag, 
in agreement with the GALEX upper limits. Using the distance modulus at 
$z=0.211$ ($m - M = 40.07$) and the photometry in Table \ref{tab:phot} 
we obtain the colors $M(NUV)-M(r)\gtrsim4.5$ and $M(r)-M(J)\sim1$.  
Galaxies with such red colors and low sSFR are usually called passive or quiescent galaxies 
to distinguish them from star-forming galaxies, 
and are consistent with an ETG population selected
morphologically \citep[e.g.,][]{Ilbert2010a,Ilbert2013a}.
Indeed, GRBH 050219A lies in a region occupied by galaxies morphologically classified as 
ETGs (\citealt{Rowlands2012a}; left panel of Fig.~\ref{fig:sfr}). 
ETGs and especially elliptical galaxies are commonly associated with 
an old stellar population, and believed to be cold-gas and
dust free. However, in the past years this picture has started to change
due to the discovery 
that many ETGs contain significant amounts of gas and dust,
and are actively forming stars, 
some with SFRs comparable to typical late-type spiral galaxies 
\citep[e.g.,][]{Fukugita2004a,Serra2012,Rowlands2012a}.

\begin{figure}[tbp]
\begin{center}
\includegraphics[width=0.48\textwidth]{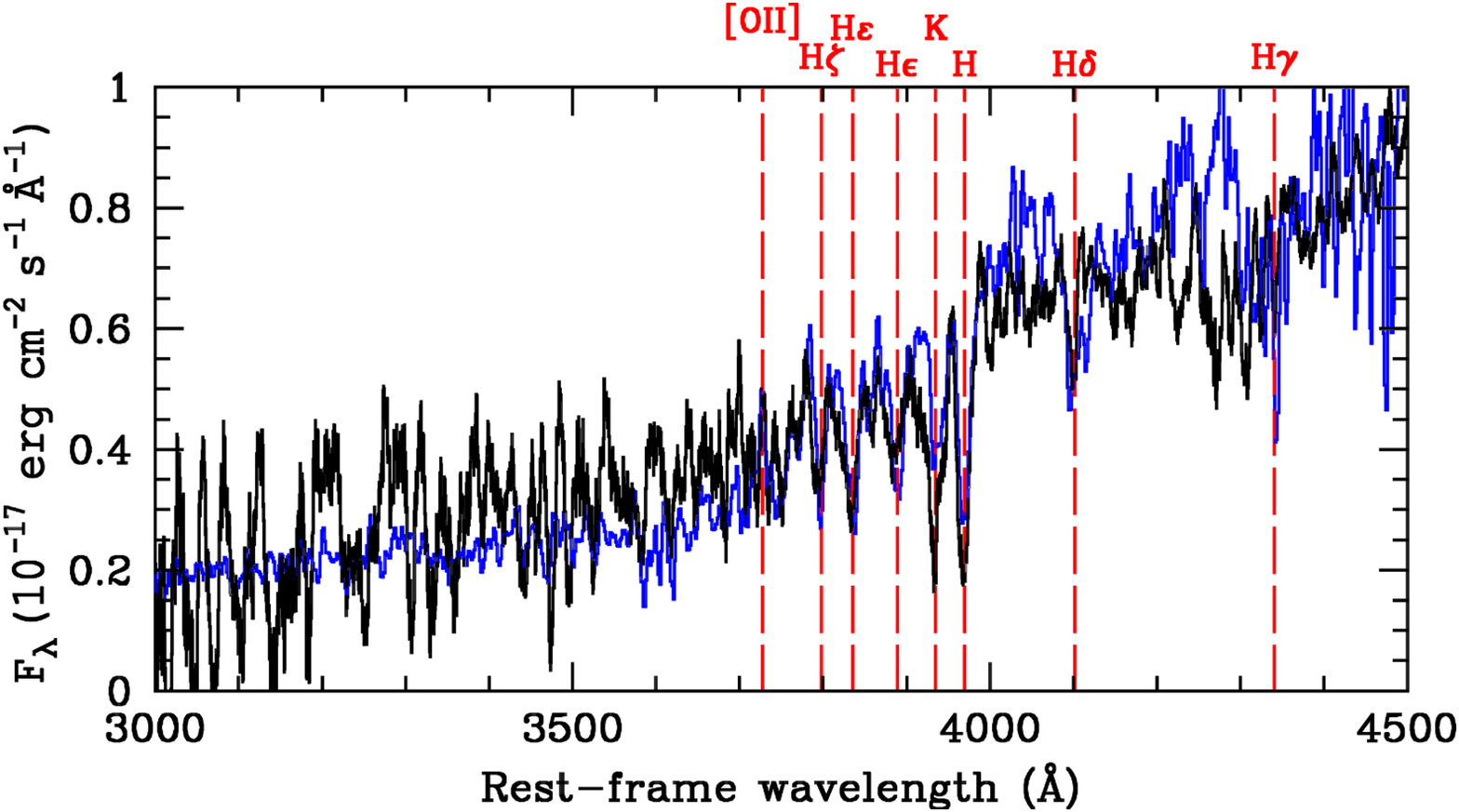}
\caption{The X-shooter spectrum of GRBH 050219A (black) compared with the spectrum of
the post-starburst galaxy GDDS-12-8139 \citep{LeBorgne2006a}.
We marked the positions of the [\ion{O}{ii}] emission line as well as the most visible absorption lines.
}
\label{fig:e+a}
\end{center}
\end{figure}

The spectrum of GRBH 050219A shows a $4000${\AA} break as well as a strong H$\delta$ absorption 
line coupled with little [\ion{O}{ii}] emission. These features are typical of
the H$\delta$-strong galaxy population 
\citep[e.g.,][]{Dressler1983a,Balogh1999a,LeBorgne2006a}.
They are post-starburst galaxies, i.e., they have undergone a recent 
break in their star-formation activity and are transiting to a quiescent phase.
They were more common at redshift $\sim1$ and have decreased to a few percent today.
In Table \ref{tab:spec} we summarise the spectral features of GRBH 050219A 
together with those of some post-starburst galaxies studied by \citet{LeBorgne2006a}. 
The host galaxy candidate of GRB 050219A
might be a low-redshift and low-mass member 
of this galaxy population. Although GRBH 050219A is less massive, its spectrum
is otherwise similar to the post-starburst galaxy GDDS-12-8139 (Fig.~\ref{fig:e+a}).
The relatively strong H$\delta$ line and the presence of [\ion{O}{ii}] emission
could be the echoes of intense episodes of star-formation that faded $\sim500$ Myr prior 
to the epoch of observation. 
An alternative explanation for the recent star-formation 
in post-starburst galaxies is the interaction of an old galaxy 
with a companion \citep[e.g.,][]{Zabludoff1996a,Goto2005a}. 
This can also be the case for 
GRBH 050219A, which possibly interacted in the past 
with the close (in projection) fuzzy source. 
We cannot prove this interaction, but we speculate 
that the small amount of star-formation responsible for the weak [\ion{O}{ii}] 
emission and the explosion of the burst may be the sign of interaction
between the quiescent galaxy and a possible companion. 

Core-collapse SNe are extremely rare in ETGs but not impossible.
For example, \citet{Kawabata2010a} discovered the Type Ib SN 2005cz in 
the local elliptical galaxy NGC 4589. Also, \citet{Perets2010a}
report that SN 2005E is hosted by an elliptical galaxy that also includes 
 a small and young stellar population with life-times of $10^7-10^8$ years, 
consistent with the core-collapse model.
\citet{Hakobyan2012a} found only four core-collapse
SNe in ETGs in a sample of 2104 SN host galaxies \citep[see also][]{Leaman2011a}.
Since these events are related to star-formation, their host galaxies
could be interacting galaxies, where a starburst was triggered by galaxy-galaxy-interaction 
\citep[e.g.,][]{Fruchter1999b,Wainwright2007,Chen2012a}.

\section{Summary \label{Sum}}

We have demonstrated that GRB 050219A is 
a rightful member of the LGRB population.
Within its XRT error circle we have 
found a relatively bright galaxy, detected from the optical to the mid-IR bands.
Its morphology and the absorption lines in the
X-shooter spectrum suggest that this is an ETG dominated by an old 
stellar population at redshift $z=0.211$.
We identify this galaxy as the most probable host galaxy of GRB 050219A
on the basis of its low chance probability. 
It has the lowest sSFR among all known LGRB hosts
($\rm{sSFR}=\rm{SFR}/\rm{M}_*\lesssim6\times10^{-12}\,{\rm yr}^{-1}$), but 
is similar to other ETGs. 
The properties of GRBH 050219A, and in particular its size and sSFR, 
are on average different from those of other hosts of 
core-collapse SNe and LGRBs.
The presence of strong stellar absorption features coupled with a weak [\ion{O}{ii}] 
emission line in the X-shooter spectrum suggest that the host galaxy candidate is a low-mass member of
the post-starburst galaxy population.
We conclude that it is the first quiescent ETG and the first post-starburst galaxy
found to probably host a LGRB.
Additional observations are required in order to 
exclude alternative host galaxies like the faint fuzzy source outside the XRT circle, 
and to study the high-mass star formation at the burst location.

Several years ago the host galaxies of LGRBs
were believed to be sub-luminous, blue compact dwarfs. 
In the past years this view has changed 
thanks to the discovery of massive and dusty hosts of dark LGRBs 
\citep[e.g.,][]{Hunt2011a,Rossi2012a,Perley2013a}.
The discovery of a LGRB 050219A in an quiescent ETG
would be further evidence that LGRBs can explode
in a variety of galaxies with the only requirement 
being an episode of high-mass star-formation. 

\begin{acknowledgements}

We are grateful to the anonymous referee for constructive comments
that improved the clarity of the manuscript. 
AR, EP and LKH acknowledge support from PRIN-INAF 2012/13. 
SP acknowledges partial support from PRIN MIUR 2009
and from ASI INAF I/004/11/1. AR and DAK acknowledge
support by the Th\"uringer Landessternwarte Tautenburg.
MJM acknowledges the support of the Science and Technology Facilities Council.
ANG acknowledges support by DFG grant Kl 766/16-1. 
S. Schulze acknowledges support from CONICYT through FONDECYT grant 3140534, from Basal-CATA PFB-06/2007, Iniciativa Cientifica Milenio grant P10-064-F (Millennium Center for Supernova Science), by Project IC120009 "Millennium Institute of Astrophysics (MAS)" of Iniciativa Cient\'ifica Milenio del Ministerio de Economi\'ia, Fomento y Turismo de Chile, with input
from "Fondo de Innovacio\'on para la Competitividad, del Ministerio de Econom\'ia, Fomento y Turismo de Chile".
PS acknowledges support through the Sofja Kovalevskaja Award from the Alexander von Humboldt Foundation of Germany.
E. Pian acknowledges partial support from contracts ASI INAF I/088/06/0, INAF PRIN 2011 and 
PRIN MIUR 2010/2011. Part of the funding for GROND (both hardware as well as personnel) was generously 
granted from the Leibniz-Prize to Prof. G. Hasinger (DFG grant HA 1850/28-1). 
We thank D. Malesani and P. M. Vreeswijk for their 
contribution to the analysis of the X-shooter spectrum, and N. 
Masetti for insightful suggestions on many aspects of this study.
This research has made use of the NASA/IPAC Extragalactic Database (NED) which is operated by
the Jet Propulsion Laboratory, California Institute of Technology, under
contract with the National Aeronautics and Space Administration. This work made
use of data supplied by the UK \swift Science Data Centre at the University of
Leicester. 

\end{acknowledgements}


\bibliographystyle{aa}
\bibliography{host050219A}



\begin{appendix} 



\section{Properties of the GRB and its afterglow \label{app:grb}}

\subsection{Properties of the prompt emission \label{app:prompt}} 

\begin{figure}[htbp]
\begin{center}
\includegraphics[width=0.48\textwidth]{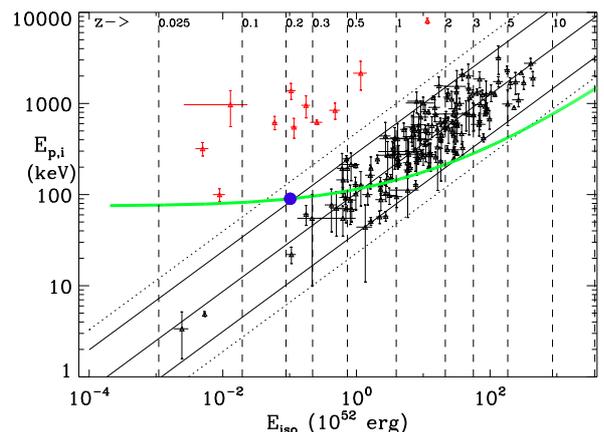}
\caption{ 
LGRBs (black) and short GRBs (red) in the ${\rm E}_{\rm p,i} - {\rm E}_{\rm iso}$ plane.
The blue dot indicates the position of GRB 050219A assuming $z = 0.211$. 
The dashed vertical lines indicate the position of the GRB at different redshifts.
The green line shows how this position changes together with the redshift.
The solid and dotted lines show the $2\sigma$ and $3\sigma$ regions respectively.
}
\label{fig:epeiso}
\end{center}
\end{figure}

\begin{figure}[htbp]
\begin{center}
\includegraphics[width=0.50\textwidth]{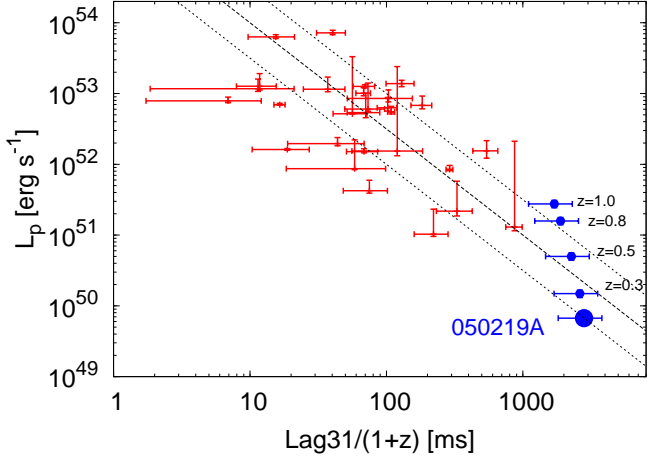}
\caption{
Isotropic peak luminosity as a function of spectral lag between BAT
channels 3 (50--100 keV) and 1 (15--25 keV). The image is adapted from \citet{Ukwatta2010a} and GRBs are marked in red.
The dotted lines indicate the estimated $1\sigma$ confidence level.
We marked in blue the position of GRB 050219A at different redshifts, including the 
 redshift of the bright galaxy at $z=0.211$ (bigger blue dot).
}
\label{fig:lag}
\end{center}
\end{figure}

In the last years, the detection of soft tails in short GRBs \citep[e.g.,  GRB 050724,][]{Barthelmy2005b} 
and of peculiar long events like GRB 060614 \citep{Gehrels2006a}, have shown the limitations of
the standard short/long classification based on duration only \citep[e.g.,][]{Zhang2009ApJ}. Thus, to
test the nature of GRB 050219A and whether it is a real LGRB, we investigate whether this burst
satisfies the ${\rm E}_{\rm p,i} - {\rm E}_{\rm iso}$ correlation \citep{Amati2006a}. 
We based our analysis on 
the spectral fits, light curves and fluences reported in \citet{Tagliaferri2005a}
and on our personal analysis of the \swift/BAT spectrum.
They show that the $\sim24$ sec pulse observed by \swift/BAT was followed by softer emission
observed with the XRT, and thus the central engine was active for at least $\sim120~\rm s$.
However, this later softer phase accounts only for $<20\%$ of the total prompt emission 
and therefore the following conclusions do not depend on
the softer emission. Luckily, the spectral peak falls within the BAT band and it is ${\rm E}_{\rm p,i}=90\pm9$ keV.
We find that assuming a redshift $z=0.211$ this burst is consistent with the 
 ${\rm E}_{\rm p,i} - {\rm E}_{\rm iso}$ correlation within $2\sigma$.
 Any value between $0.2\lesssim z \lesssim2$
 is good within $2\sigma$. The GRB would lie in the region occupied by short GRBs
 only for $z\lesssim0.1$ (see Fig.~\ref{fig:epeiso}).

We also checked the lag-luminosity relation, following the method described in \citet{Ukwatta2010a}.
We assumed the same redshift as above, 
and we measure a peak luminosity of $(6.7\pm0.8)\times 10^{49}\,$erg$\,$s$^{-1}$.
We find a lag of $3.4\pm1.2$ s between BAT channels 3 (50--100 keV) and 1 (15--25 keV)
and of  $1.4\pm0.6$ s between BAT channels 4 (100--200 keV) and 2 (25--50 keV).
Figure~\ref{fig:lag} shows that GRB 050219A has one of the largest lags measured  
for a LGRB and it is in very good agreement
with the lag-luminosity relation for LGRBs presented in \citet{Ukwatta2010a}.

Therefore, even though a redshift of about $0.5-1$ 
would place it in the regions of the  ${\rm E}_{\rm p,i} - {\rm E}_{\rm iso}$ and 
lag - luminosity planes most populated by typical LGRBs (see Figs.~\ref{fig:epeiso} and \ref{fig:lag}),
 the energetics, luminosity, spectrum and timing properties of GRB 050219A
 are consistent with the hypothesis that it is a LGRB at the
 redshift of the putative host galaxy.

\subsection{GRB 050219A was a dark burst \label{app:dark}}


Within the standard afterglow theory \citep{Sari1998a} the X-ray afterglow should
have an X-ray spectral slope $\beta_{X}\gtrsim0.5$, and $\beta_{X}\gtrsim1$ in case of a break below the X-ray frequency
which is usually the case \citep{Greiner2011}.
 \citet{Jakobsson2004a} quantify the optical dimness by testing whether 
the optical to X-ray spectral slope $\beta_{OX}$ is lower than $0.5$.
We downloaded and analysed the XRT spectrum (both pc and wt modes) and we found $0.6<\beta_X<1.2$, consistent with 
the values in the \swift/XRT on-line repository,
but not good enough to distinguish between the two scenarios. However, without a break between optical
and X-rays (i.e., $\beta_{OX}\gtrsim1$), 
the afterglow would have had $R\gtrsim$18 at the time of the
MOA observations, surely dominating over the galaxy light and well-detected, but we do not see this in our re-analysis of the data.
In the case of a break between optical and X-rays, lying close to the X-ray band, the afterglow should have been $R\sim20$ or 
brighter. This is similar to the MOA upper limit, 
but because of the bad seeing it could be easily confused with the host with $R\sim20$.
Thus, we conclude that the burst had a $\beta_{OX}\lesssim0.5$ and therefore it can be considered a marginal member of the dark GRB population
according to the \citet{Jakobsson2004a} criterion.



\section{Photometry \label{app:phot}}


\begin{figure}[tp!]
\begin{center}
\includegraphics[width=0.45\textwidth]{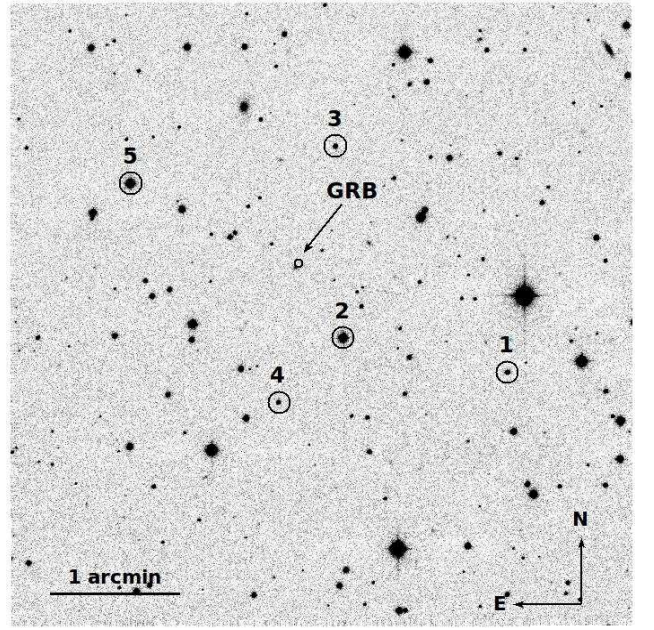}
\caption{ 
Finding chart of the field of GRB 050219A (GROND $r^{\prime}$-band). 
The X-ray afterglow position and the secondary
photometric standards used (Table~\ref{tab:std}) are indicated. 
}
\label{fig:finding}
\end{center}
\end{figure}

\begin{table*}[tbp]
\begin{minipage}[t]{\textwidth}
\normalsize\addtolength{\tabcolsep}{+5pt}
\renewcommand{\footnoterule}{}
\centering
\caption{Observation log for the GRB 050219A field.}
\begin{tabular}{lccccccc}
\toprule
 Filter & Wavelength & Telesc./Instr. & Magnitudes      & AB Magnitudes & Flux density   & upper       & lower	  \\
        & ($\mu m$)    &                & (Observed)      &   (Ext. cor.)  & ($\mu Jy$)  & ($\mu Jy$)	& ($\mu Jy$)  \\
\midrule             
$FUV$          & 0.152  & GALEX        & $>21.6$         & $>20.5$ &  $<23.89$ &	  --  &  -- 	\\
$NUV$          & 0.193  & GALEX        & $>21.7$	      & $>20.6$ &  $<20.99$ &	  --  &  --   	     \\
$uvw2$         & 0.225  & \swift/UVOT  & $>21.3$	      & $>21.8$	&  $<7.19$  &	  --  &  --   	\\
$uvm2$         & 0.227  & \swift/UVOT  & $>20.5$	      & $>20.7$	&  $<18.86$ &	  --  &  --   	\\
$uvw1$         & 0.260  & \swift/UVOT  & $>21.3$	      & $>21.7$	&  $<7.42$  &	  --  &  --   	\\
 $u$           & 0.347  & \swift/UVOT  & $>21.4$	      & $>21.6$	&  $<8.56$  &	  --  &  --   	\\
 $b$           & 0.430  & \swift/UVOT  & $>21.7$	      & $>20.8$	&  $<16.90$ &	  --  &  --   	\\
 $v$           & 0.543  & \swift/UVOT  & 20.82 $\pm$ 0.12& 20.31	& 27.86 &  31.12  &   24.95		 \\
 $g^\prime$    & 0.459  & 2.2m/GROND   & 21.49 $\pm$ 0.04& 20.85	& 16.96 &  17.64  &   16.38  \\  
 $r^\prime$    & 0.622  & 2.2m/GROND   & 20.17 $\pm$ 0.04& 19.76	& 46.20 &  47.72  &   44.74  \\  
 $i^\prime$    & 0.764  & 2.2m/GROND   & 19.70 $\pm$ 0.04& 19.38	& 65.55 &  67.89  &   63.30  \\ 
 $z^\prime$    & 0.899  & 2.2m/GROND   & 19.41 $\pm$ 0.04& 19.18	& 78.77 &  81.58  &   76.06		\\	 
 $J$           & 1.240  & 2.2m/GROND   & 17.74 $\pm$ 0.04& 18.52	& 145.06&  149.82 &   140.46 \\  
 $H$           & 1.647  & 2.2m/GROND   & 17.01 $\pm$ 0.10& 18.30	& 178.52&  195.74 &   162.81 \\   
 $K_s$         & 2.170  & 2.2m/GROND   & 16.66 $\pm$ 0.12& 18.44	& 156.30&  174.56 &   139.94		\\  
 $R$           & 0.660  & VLT/FORS2   & 20.2  $\pm$ 0.1 & 20.0	& 37.19 &  33.91  &   40.78 \\
 $3.6\mu m$     & 3.543  & {\it Spitzer}/IRAC & 19.10 $\pm$ 0.02& 19.10    & 85.00 &  87.00  &   83.00   \\
 $5.8\mu m$     & 5.711  & {\it Spitzer}/IRAC & 19.5  $\pm$ 0.2 & 19.5    & 58.00 &  66.00  &   50.00   \\
 $24\mu m$      & 23.68  & {\it Spitzer}/MIPS & 20.7  $\pm$ 0.3 & 20.7    & 19.00 &  24.00  &   14.00			   \\
100     & 100       & {\it Herschel}/PACS      & $>14.6$	      & $>14.6$   & $<5400$	   &	 --      &  	 --  \\
160     & 160       & {\it Herschel}/PACS      & $>14.4$	      & $>14.4$   & $<15600$	   &	 --      &  	 --  \\
250     & 250       & {\it Herschel}/SPIRE     & $>13.4$	      & $>13.4$   & $<16500$	   &	 --      &  	 --  \\
350     & 350       & {\it Herschel}/SPIRE     & $>13.1$	      & $>13.1$   & $<21000$	   &	 --      &  	 --  \\
500     & 500       & {\it Herschel}/SPIRE     & $>12.9$	      & $>12.9$   & $<24600$	   &	 --    &  	 -- \\
\bottomrule							     	     
\end{tabular}
\tablefoot{                   
Column 4 gives observed magnitudes and their errors in the native system of the filters. 
Column 5 has AB magnitudes corrected for Galactic extinction ($E(B-V)=0.16$ mag). 
The last 3 columns are the corresponding flux densities and their 
upper and lower value in $\mu Jy$ after correction for 
Galactic extinction. The upper
limits are 3$\sigma$ above the background. To convert UVOT magnitudes into the AB system 
we used the following conversion: $uvw2=+1.73$, $uvm2=+1.69$, $uvw1=+1.51$,
$u=+1.02$, $b=+0.13$, $v=+-0.01$. For the FORS2/$R$-band we used $R=+0.23$.
}
\label{tab:phot}
\end{minipage}
\end{table*}

\begin{table*}[tbp]
\begin{minipage}[t]{\textwidth}
\renewcommand{\footnoterule}{}
\centering
\caption{Secondary standard stars within 4 arcmin of the afterglow position
(Fig.~\ref{fig:finding}).}
\begin{tabular}{r|c|ccccccc}
\toprule 
\# & R.A., Dec. (J2000) &  $g^\prime $  &  $r^\prime $  &  $i^\prime $ &  $z^\prime $  &  $J$  & $H$  &  $K_s$ \\
\midrule
1 &11:05:30.50 $-$40:41:53.7&	19.03(1)&  18.06(1)&  17.70(2)& 17.53(1) &16.19(11) &  15.50(12) &  15.58(24)   \\
2 &11:05:37.16 $-$40:41:37.7&	16.27(1)&  15.23(1)&  14.89(2)& 14.67(1) &13.61(3) &  13.09(3) &  12.95(3)   \\
3 &11:05:37.48 $-$40:40:08.1&	18.97(1)&  17.90(1)&  17.55(2)& 17.32(1) &16.21(12) &  15.54(14) &  15.61(21)   \\
4 &11:05:39.79 $-$40:42:07.8&	19.75(2)&  18.38(1)&  17.87(2)& 17.58(1) &16.36(12) &  15.88(18) &  15.39(19)   \\
5 &11:05:45.80 $-$40:40:25.5&	15.95(1)&  15.17(1)&  14.93(2)& 14.74(1) &13.77(2) &  13.36(03) &  13.21(3)   \\

\bottomrule
\end{tabular}
\tablefoot{ Numbers in parentheses give the
photometric $1\sigma$ statistical uncertainty of the secondary standards 
in units of ten milli-mag.
}
\label{tab:std}
\end{minipage}
\end{table*}

\end{appendix}


\end{document}